\documentclass[12pt]{article}

\newcommand{\citep}{\cite}
\usepackage{amsfonts}
\usepackage{amssymb}
\usepackage{epsfig}
\usepackage{rotating}
\usepackage{lineno}
\usepackage{color}
\usepackage{amsmath}
\usepackage{mathrsfs}
\usepackage{hyperref}
\usepackage{setspace}
\usepackage{grffile}
\usepackage{multirow}
\usepackage{verbatim}   
\usepackage{url}
\usepackage{bm}

\makeatletter
\def\url@leostyle{%
  \@ifundefined{selectfont}{\def\UrlFont{\sf}}{\def\UrlFont{\small\ttfamily}}}
\makeatother
\usepackage{hyperref}
\hypersetup{pagebackref, bookmarks=true,
citecolor=blue, urlcolor=blue}
\let\epsilon=\varepsilon

\begin{document}


\title{Spatio-temporal patterns of influenza B proportions}

\author{
 Daihai He$^1$, Alice PY Chiu$^1$, Qianying Lin$^1$ and Duo Yu $^{1,2}$
\vspace{0.2cm}\\
{\footnotesize $^1$ Department of Applied Mathematics, Hong Kong Polytechnic University, Hung Hom, Kowloon,}\\
{\footnotesize Hong Kong (SAR) China}\\
{\footnotesize $^2$ School of Mathematics and Information Science, Shaanxi Normal University, Xi'an Shaanxi, P.R. China}\\
\\
}

\maketitle

\section*{Abstract}
We study the spatio-temporal patterns of the proportion of influenza B out of laboratory confirmations of both influenza A and B, with data from 139 countries and regions downloaded from the FluNet compiled by the World Health Organization, from January 2006 to October 2015, excluding 2009. We restricted our analysis to 34 countries that reported more than 2000  confirmations for each of types A and B over the study period. We find that Pearson's correlation is 0.669 between effective distance from Mexico and influenza B proportion among the countries from January 2006 to October 2015. In the United States, influenza B proportion in the pre-pandemic period (2003-2008) negatively correlated with that in the post-pandemic era (2010-2015) at the regional level. Our study limitations are the country-level variations in both surveillance methods and testing policies. Influenza B proportion displayed wide variations over the study period. Our findings suggest that even after excluding 2009's data, the influenza pandemic still has an evident impact on the relative burden of the two influenza types. Future studies could examine whether there are other additional factors. This study has potential implications in prioritizing public health control measures.
\section{Introduction}\label{S:intro}

Globally, influenza is an important cause of morbidity, mortality and hospitalization. Two major types circulate among human population, type A and type B, and they shared many common morphological and epidemiological characteristics. They also have similar clinical presentations \cite{Osterhaus}. However, they differ in the animal reservoirs that they reside, influenza A virus infects mainly the  mammalian species and birds, while influenza B virus infects only humans and seals. Influenza B virus displayed less (at slower rate) antigenic drift and can cause significant epidemics but not pandemic, unlike influenza A \cite{Ferguson}.
Previous studies have investigated into the trends and burden of influenza B \cite{Harvala,PaulChan,JYWong}. Glezen et al. suggested that there was an increasing trend of the influenza B proportion in both the United States (US) and Europe between 1994 and 2011 \cite{Glezen}. Other studies have also explored the potential roles of air travel on the global and regional spread of influenza. These studies showed that air travel volumes and flight distance are important factors driving the spread of influenza \cite{Li, Grais2003, Grais2004, Ruan}. However, to our knowledge, there is a lack of recent literature that examined the global patterns of influenza B proportion after the 2009 influenza pandemic.
	
The aim of our study is to examine the spatio-temporal patterns of influenza B proportion in the post-pandemic era at both global and regional level. We focus on weekly laboratory confirmations from Hong Kong Special Administrative Region (SAR) China and other 73 countries that have the most laboratory-confirmations of influenza A and influenza B from 2006 to 2015. We exclude the year of 2009 to avoid the impact of widely different testing efforts (i.e. testing policies or testing practices) among countries during the 2009 A(H1N1) influenza pandemic, which was originated from Mexico \cite{WHO2009}. We define the influenza B proportion as the number of laboratory-confirmations of influenza B out of the total of influenza types A and B confirmations over the whole study period. This measure reflects the relative health care burden of influenza B out of both influenza A and B cases. We focus on two research hypotheses:

(1) There is a linear association between influenza B proportion and effective distance from Mexico, given that the pandemic H1N1 strain spread from Mexico and Northern America to the rest of the world. Simonsen et al. \cite{Simo+13} found ``far greater pandemic severity in the Americas than in Australia, New Zealand, and Europe'' in 2009. He et al. \cite{He+15} found that the H1N1pdm (pandemic strain) skipped a large part of Europe and East Asia, but not Northern America.

(2) In the US, there is a linear association between the pre-pandemic and post-pandemic era at the regional level, given that influenza B and influenza A (especially the H3N2 strain) showed anti-phase patterns. Namely, when influenza A/H3N2 are severe, the other strains are mild.

\subsection*{Data Collection}
Influenza data are downloaded from the FluNet, a publicly available and real-time global database for influenza virological surveillance compiled by the World Health Organization. We obtained data from January 2006 to October 2015, covering 138 countries. We excluded data from the year 2009, to avoid the impact due to excessive testing in many countries. We then summed up the total confirmations of influenza A and influenza B separately, and computed the proportions of influenza B, i.e. $r_i=\sum\limits_t B_t /(\sum\limits_t A_t+\sum\limits_t B_t)$, where $A_t$ and $B_t$ are weekly confirmations over the period. We restricted to those countries where $\sum\limits_t A_t>\theta$ and $\sum\limits_t B_t>\theta$ to reduce the impact of statistical noise created by small numbers from small populations on this ratio. We considered $\theta=500$ and $\theta=2000$.

We also downloaded laboratory confirmations for influenza A and influenza B from the Center of Health Protection in Hong Kong from January 1998 to October 2015 (www.chp.gov.hk), FluView from the Centers for Disease Control and Prevention in the US from January 1997 to October 2015 (www.cdc.gov/flu/weekly/). Again, we computed the influenza B proportion using the same method for the period 2006-2015. We also conduct regional level study on the correlation between influenza B proportion before and after the 2009 influenza pandemic in the US.

In addition, we collected information about individual countries including their population size as of 2005 \cite{UnitedNations}, longitudinal and latitudinal information \cite{ThematicMapping} and flight data from the Official Airline Guide (OAG, http://www.oag.com).

\subsection*{Methods}

For hypothesis 1, we computed the Pearson's correlation between influenza B proportion and the effective distance from Mexico. As a comparison, we repeat the analysis using China as the reference country, instead of Mexico. Besides, we also computed the Pearson's correlation between influenza B proportion and the longitude and latitude (absolute value). We varied the threshold of influenza specimen at 500 and 2000. We also considered different time periods, i.e. pre-pandemic period, post-pandemic period or the entire study period covering both.

For hypothesis 2, we computed the Pearson's correlation between influenza B proportion in the pre-pandemic and post-pandemic era among the ten census region in the United States. We defined pre-pandemic period with varying start year between 1997 and 2007, and the end year as 2008. The post-pandemic period is defined as January 2010 to October 2015. We computed the Pearson's correlation for each of these start years.

Also, we constructed four different statistical models to determine how the influenza B proportion is associated with the other factors. Model 1 is a linear mixed effect model. The response factor is influenza B proportion. The independent factors include population size, longitude, absolute latitude, effective distance as fixed factors, and geographic region as a random factor.

Model 2 is also a linear mixed effect model, the factors are the same as Model 1, except that we use the number of laboratory specimens tested in place of the population size.

Model 3 is a linear model. The response factor is influenza B proportion. The independent factors include population size, longitude, absolute latitude and effective distance.

Model 4 is also a linear model. We have the same factors as in Model 3, except that we use the number of laboratory specimens tested in place of the population size.

In addition, we create a linear model (Model 5), with influenza B proportion as response factor, and population size, longitude, absolute latitude, effective distance and region as independent factors. We remove one independent factor at a time, and then re-assess the model fit by re-running the linear model and assessing its Akaike Information Criterion (AIC).

All statistical analyses are conducted using statistical package R version 3.2.2. Statistical significance is assessed at 0.05 level.

\subsection*{Effective Distance}

The global air traffic data from the Official Airline Guide (OAG, http://www.oag.com) were used to calculate the effective distance. These data provide the number of seats on each flight route between pairs of worldwide airports in 2009. We built the global air network at the country level by aggregating each group of airports that are located in the same country into a single node. There are in total 209 countries (nodes) and more than 4,700 connections (edges) among different countries.

As in \cite{BrocHelb13}, the effective distance from node $i$ to one of its directly connected node $j$ is measured by $D_{ij}=1-\ln(P_{ij})$, where $P_{ij}=T_{ij}/T_{i}$ is the relative mobility rate from $i$ to $j$. $T_{ij}$ represents the number of passengers from node $i$ to $j$, and $T_i=\sum_{j \in v(i)}T_{ij}$  represents the total number of passengers from node $i$. The effective distance from seed node to another node in the network is defined by the distance of the shortest path. For any path between the two nodes, its distance is the sum of effective distance for every edge on that path.

The definition of effective distance seems quite arbitrary. Some alterative choices can be used for comparison. For example, define $D_{ij}=-\ln(\omega_{ij})$, where $\omega_{ij}=T_{ij}/N_i$ and $N_i$ is the population size of country $i$. These two effective distances are highly correlated ($p-$value $< 0.001$).

We found that the influenza B proportion from 74 localities are evidently correlated with their effective distance of from Mexico, using either definition of effective distance.

\section{Results}

\subsection{Global Patterns}
\begin{figure}[ht!]
\centerline{\includegraphics[width=20cm]{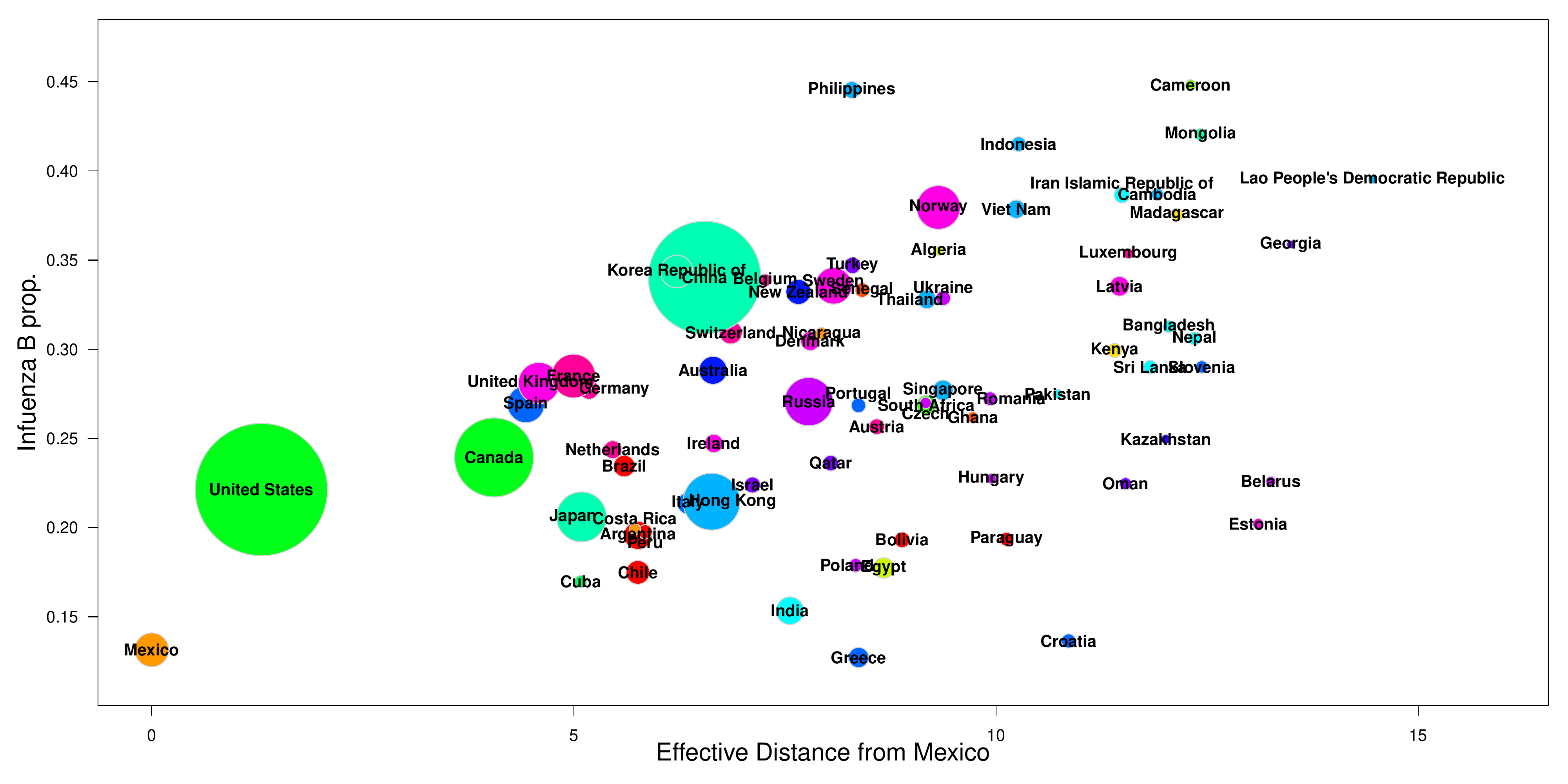}}
\caption{Influenza B proportion from 73 countries against their effective distance from Mexico. We use disks to represent individual countries. The size of the disk is proportional to the square root of the original size in order to enlarge the smaller values. Countries are coloured according to their geographic region or sub-region.}
\label{Fig:effd}
\end{figure}

Figure~\ref{Fig:effd} shows the influenza B proportion versus the effective distances from 73 localities to Mexico. Influenza B proportion is significantly and positively correlated with the effective distance, i.e., the further from Mexico the higher the proportion. Pearson's correlation is 0.433 (95\%CI: 0.225, 0.603) and  $p-$value $<0.001$. The correlation is 0.669, (95\% CI: 0.422,0.823, $p-$value $<0.0001$) among the top 33 countries where both types A and B confirmations are more than 2000. It is also evident that countries in the same geographic region (i.e. as represented by the same colour) tend to have similar influenza B proportions.

As a comparison, if we choose China as the reference country, the correlation of influenza B proportion versus the effective distance from 73 localities to China is not statistically significant with $p-$value = 0.379.

\begin{table}[h!]
\begin{center}
\caption{Summary of all Pearson's correlation tests. $\rho$ stands for correlation coefficient, and $\theta$ stands for the threshold of influenza specimens.}

\begin{tabular}{rrrrrrrrr}
\hline
factor& ref. country& time interval& $d.f.$& $\theta$& $\rho$& 95\%CI& $p$-value& signif\\
effd1& Mexico& 2006-1 to 2015-40& 71& 500& 0.433& (0.225, 0.603)& 0.000131& ***\\
effd2& Mexico& 2006-1 to 2015-40& 71& 500& 0.474& (0.273, 0.635)& 2.3e-05& ****\\
effd1& China& 2006-1 to 2015-40& 71& 500& -0.104& (-0.327, 0.129)& 0.379363& \\
longitude& -& 2006-1 to 2015-40& 72& 500& 0.445& (0.241, 0.611)& 7.1e-05& ****\\
latitude& -& 2006-1 to 2015-40& 72& 500& 0.014& (-0.215, 0.242)& 0.905432& \\
\hline
effd1& Mexico& 2006-1 to 2008-52& 29& 2000& 0.192& (-0.174, 0.511)& 0.30128& \\
effd2& Mexico& 2006-1 to 2008-52& 29& 2000& 0.235& (-0.13, 0.544)& 0.20309& \\
effd1& China& 2006-1 to 2008-52& 29& 2000& -0.032& (-0.382, 0.326)& 0.864635& \\
longitude& -& 2006-1 to 2008-52& 29& 2000& 0.029& (-0.329, 0.379)& 0.878559& \\
latitude& -& 2006-1 to 2008-52& 29& 2000& 0.236& (-0.129, 0.545)& 0.200923& \\
\hline
effd1& Mexico& 2010-1 to 2015-40& 31& 2000& 0.677& (0.435, 0.828)& 1.5e-05& ****\\
effd2& Mexico& 2010-1 to 2015-40& 31& 2000& 0.677& (0.434, 0.828)& 1.5e-05& ****\\
effd1& China& 2010-1 to 2015-40& 31& 2000& -0.068& (-0.402, 0.282)& 0.706725& \\
longitude& -& 2010-1 to 2015-40& 31& 2000& 0.515& (0.209, 0.729)& 0.002158& **\\
latitude& -& 2010-1 to 2015-40& 31& 2000& 0.099& (-0.253, 0.428)& 0.581985& \\
\hline
effd1& Mexico& 2006-1 to 2014-52& 31& 2000& 0.676& (0.433, 0.827)& 1.6e-05& ****\\
effd2& Mexico& 2006-1 to 2014-52& 31& 2000& 0.676& (0.434, 0.828)& 1.5e-05& ****\\
effd1& China& 2006-1 to 2014-52& 31& 2000& -0.175& (-0.489, 0.18)& 0.33128& \\
longitude& -& 2006-1 to 2014-52& 31& 2000& 0.462& (0.141, 0.695)& 0.006846& **\\
latitude& -& 2006-1 to 2014-52& 31& 2000& 0.164& (-0.19, 0.48)& 0.361159& \\
\hline
effd1& Mexico& 2006-1 to 2015-40& 31& 2000& 0.669& (0.422, 0.823)& 2.1e-05& ****\\
effd2& Mexico& 2006-1 to 2015-40& 31& 2000& 0.673& (0.429, 0.826)& 1.8e-05& ****\\
effd1& China& 2006-1 to 2015-40& 31& 2000& -0.083& (-0.415, 0.268)& 0.645017& \\
longitude& -& 2006-1 to 2015-40& 31& 2000& 0.495& (0.182, 0.716)& 0.003429& **\\
latitude& -& 2006-1 to 2015-40& 31& 2000& 0.119& (-0.234, 0.444)& 0.5106& \\
\hline
\end{tabular}

\end{center}
\label{Table1}
\end{table}

Table 1 shows the results of the Pearson's correlation tests of the effective distance from Mexico or China, longitude and absolute latitude. We also varied the threshold levels to 500 or 2000, and the time period to include the pre-pandemic period only, post-pandemic period only or the entire study period covering both. For effective distance from Mexico, there is a strong positive correlation of 0.669 (95\% CI: 0.422, 0.823) (using definition 1)  and 0.673 (95\% CI: 0.429, 0.826) (using definition 2) demonstrated when we consider a threshold of 2000 for the entire study period (p-value $< 0.0001$ in both cases). There was a weak negative correlation for the effective distance from China but they were not statistically significant. For longitude, medium correlations of 0.495 (95\% CI: 0.182, 0.716) and 0.445 (95\% CI: 0.241, 0.611) were demonstrated when we consider the entire study period and using thresholds of 2000 and 500 respectively. For absolute latitude, there was a weak positive correlation in general but none of them has reached statistical significance at levels 0.05.

In the supplementary material, our statistical models (Models 1-4) also demonstrated consistent results. In all four models, effective distance was the strongest predictor, while longitude was the second strongest predictor. In Models 3 and 4, both effective distance and longitude were significant at the 0.05 level in these linear models.
In Model 5, we assessed the effect of single term deletions from a full linear model. Among the five independent factors being studied, i.e. population size, longitude, absolute latitude, effective distance and region, we found that effective distance (Mexico) shows the strongest improvement.
Here the smaller the absolute value AICc is, the better the model is. The results are consistent with the other models, that the effective distance is the most important factor.

\begin{figure}[ht!]
\centerline{\includegraphics[width=20cm]{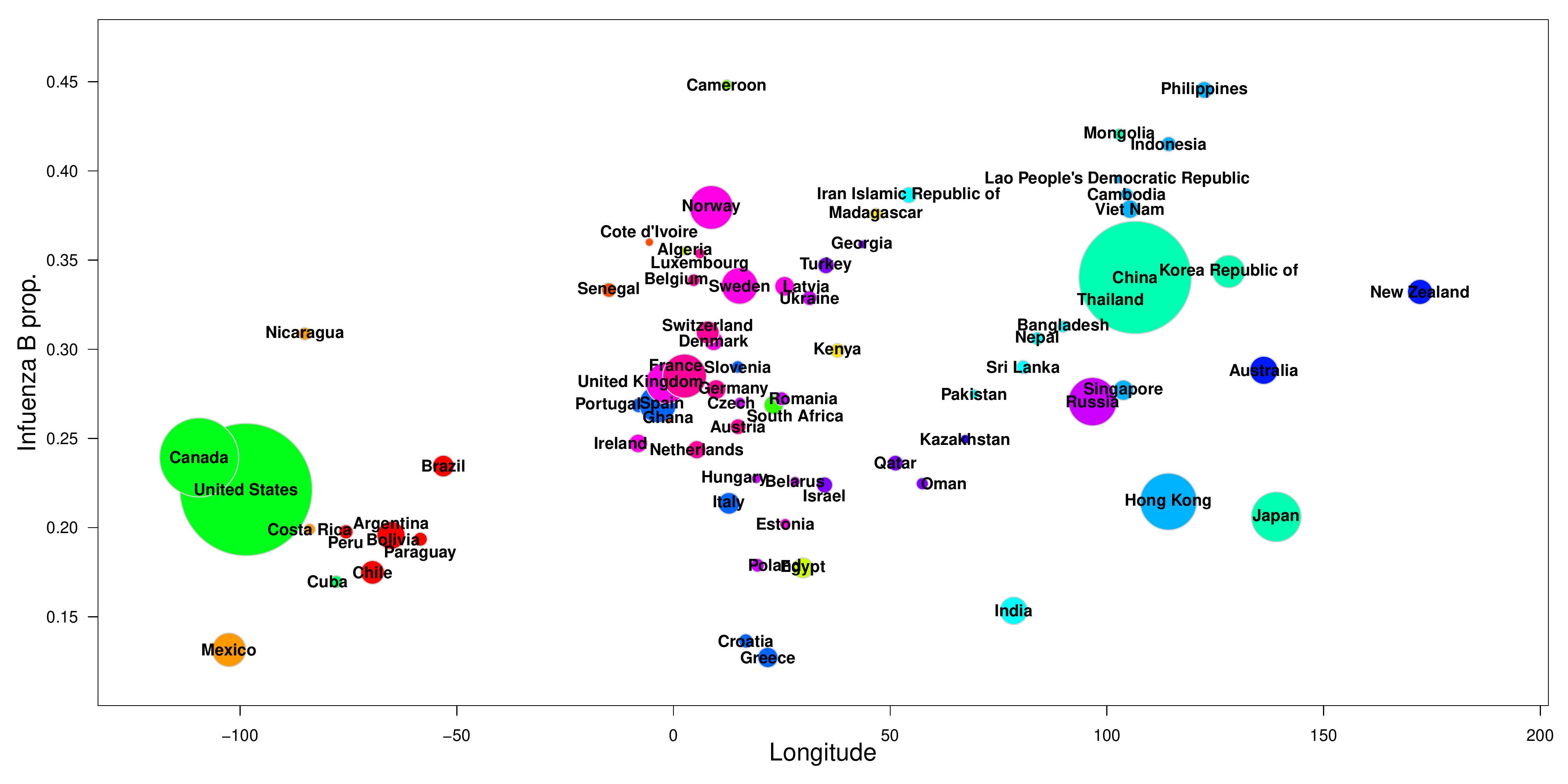}}
\caption{Influenza B proportion from 74 localities against their longitudes. We use disks to represent individual countries. The size of the disk is proportional to the square root of the total confirmations of influenza A and influenza B. Countries are coloured according to their geographic region or sub-region.}
\label{Fig:longitude}
\end{figure}

The significantly positive correlation we found here continues to hold if we restrict to the post-pandemic era (2010-2015) and excluding the pre-pandemic era (2006-2008). However, the pre-pandemic era (2006-2008) alone is not statistically significant.

As a comparison, the ratio between the two influenza A subtypes H1N1 and H3N2 is not statistically significant. These results suggest that the significantly different virus evolutionary rate between influenza B and influenza A could have played a role in the patterns of influenza B proportion.

\begin{figure}[ht!]
\centerline{\includegraphics[width=20cm]{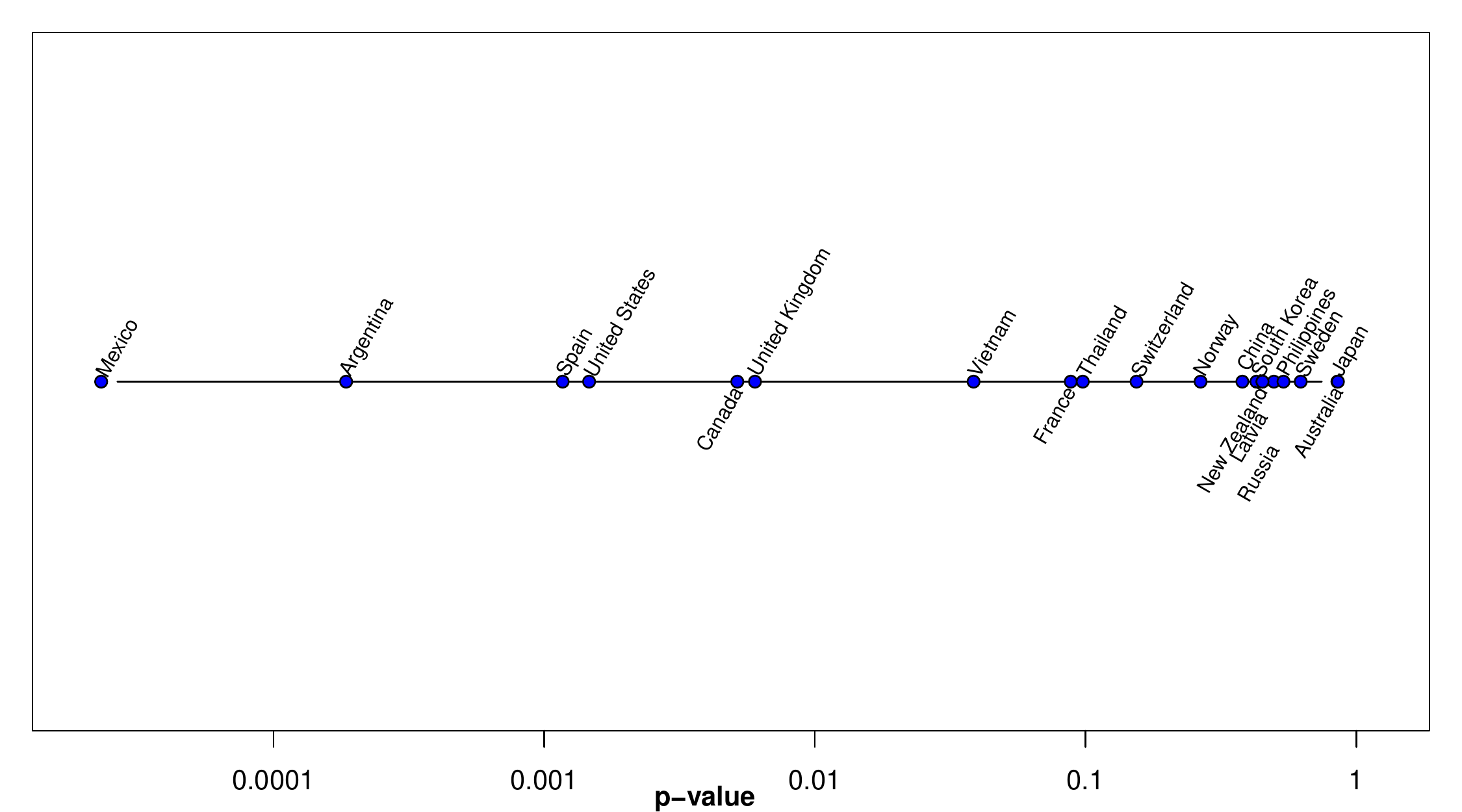}}
\caption{The $p-$values of correlations between influenza B proportion and the effective distance from a reference country.}
\label{Fig:pvalue}
\end{figure}

We show the results with either Mexico or China as the reference in Table 1. We also checked the scenarios with other countries with large influenza reports as the reference in the calculation of effective distance. We show the $p-$values in Figure~\ref{Fig:pvalue}. It can be seen that Mexico attains the smallest $p-$value. Interestingly, the other two Spanish speaking countries, Argentine and Spain, also have relatively small $p-$values when they are as the reference countries.

Figure~\ref{Fig:longitude} show that the influenza B proportion also positively correlated with the longitude of countries (Pearson's correlation=0.445,  (95\%CI: 0.241, 0.611), p-value $< 0.0001$). Even though the longitude was not a suitable factor, because, given than the world takes a global shape, where longitude of $-180 \equiv +180$. These differences among Northern America, Europe and Asia are evident and striking.

To our knowledge, patterns found in Figures \ref{Fig:effd} and \ref{Fig:longitude} are novel. These patterns echo previous studies\cite{Hinds,ECDC-WHO,CDC,Simo+13}, which reported different severity of influenza A (H1N1pdm) or influenza B between Europe and North America.

Note that the global surveillance effort in the post-pandemic era in terms of the number of specimens processed in most countries and the numbers of countries reporting data to FluNet has been improved after the 2009 pandemic. In the United States, where surveillance has been relatively consistent starting from 1997, we found an interesting negative correlation of the influenza B proportion in the post 2009 pandemic era versus the pre-2009 pandemic period. We show the results in the following section.

\subsection{Regional patterns in the United States}

\begin{figure}[ht!]
\centerline{\includegraphics[width=18cm]{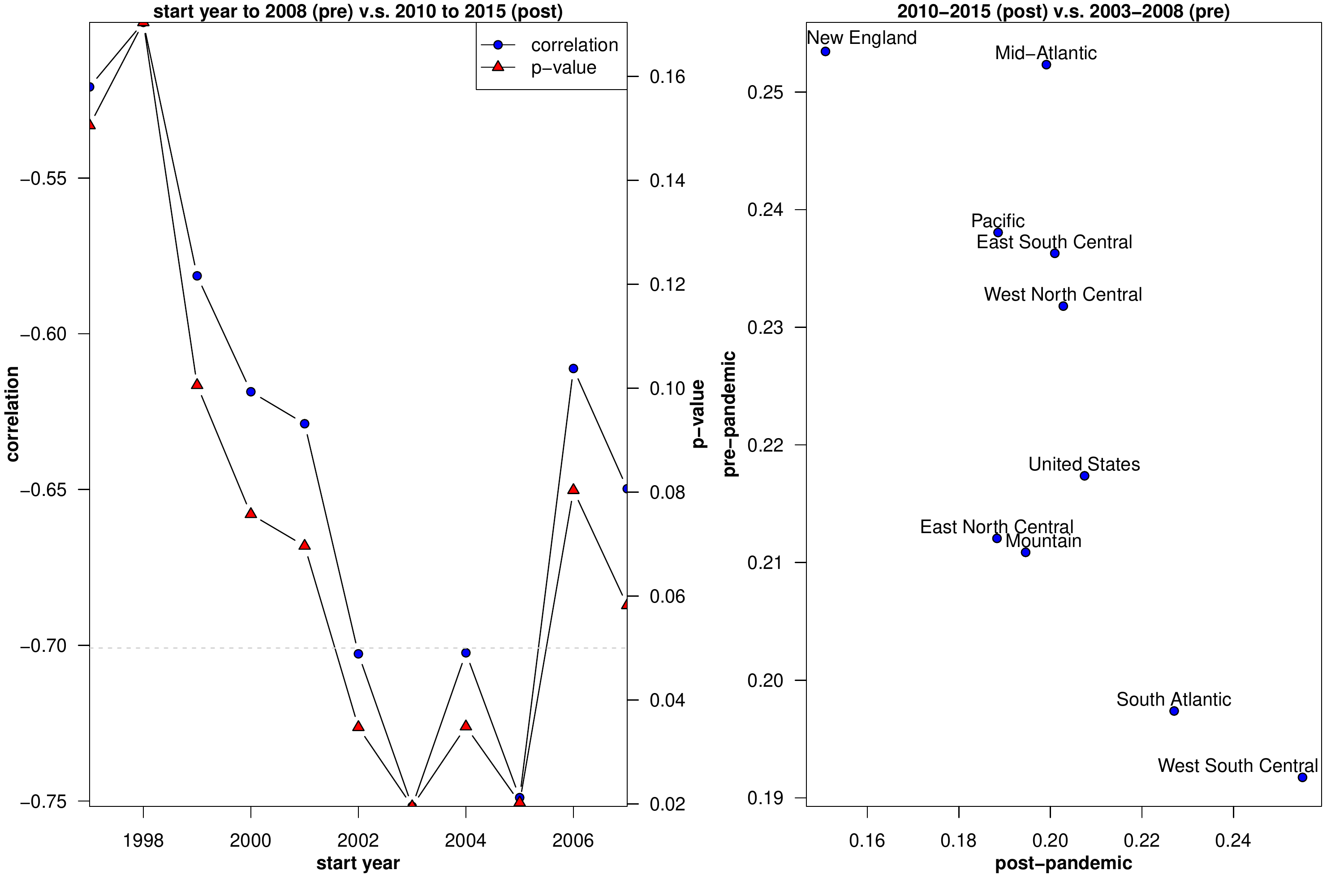}}
\caption{Influenza B proportion in the pre-pandemic period negatively correlates with that in the post-pandemic era among ten census regions of the United States. The post-pandemic era is defined from Jan 2010 to August 2015. We vary the start of the pre-pandemic from 1997 to 2007, while fix the end of the pre-pandemic era is 2008. It is clear that there is a window period such that the influenza B proportion shows a significant negative correlation between pre- and post-pandemic.  (a) correlation and p-value as functions of start year of the pre-pandemic period. The x-axis show the start of the pre-pandemic period. (b) the correlation between 2010-2015 versus 2003-2008.}
\label{Fig:US}
\end{figure}

Figure~\ref{Fig:US} shows the correlation of influenza B proportion in the pre-pandemic period versus that in the post-pandemic era among ten census regions of the US. We define the post-pandemic era as from Jan 2010 to August 2015. We vary the start of the pre-pandemic from 1997 to 2007, while we fix the end of the pre-pandemic era to be 2008. It is clear that there is a window period such that the influenza B proportion shows a significantly negative correlation between pre- and post-pandemic era, and results are shown in Table 2. The Pearson's correlation between the influenza B proportions in pre-pandemic 2003-2007 and those in post-pandemic 2010-2015 is -0.752, 95\% CI is (-0.944, -0.175), and the $p$-value is 0.0195.

\begin{table}[h!]
\caption{Summary of all Pearson's correlation tests for the US. $\rho$ stands for correlation coefficient.}
\begin{center}
\begin{tabular}{crrrr}
\hline
time interval& $\rho$& 95\% CI& $p$-value& signif\\
1997-2008 vs 2010-2015& -0.5208& (-0.88, 0.219)& 0.1506& \\
1998-2008 vs 2010-2015& -0.5001& (-0.874, 0.246)& 0.1704& \\
1999-2008 vs 2010-2015& -0.5814& (-0.899, 0.135)& 0.1006& \\
2000-2008 vs 2010-2015& -0.6186& (-0.909, 0.077)& 0.0757& \\
2001-2008 vs 2010-2015& -0.6289& (-0.912, 0.061)& 0.0696& \\
2002-2008 vs 2010-2015& -0.7027& (-0.932, -0.072)& 0.0347& *\\
2003-2008 vs 2010-2015& -0.7517& (-0.944, -0.175)& 0.0195& *\\
2004-2008 vs 2010-2015& -0.7024& (-0.932, -0.072)& 0.0349& *\\
2005-2008 vs 2010-2015& -0.7489& (-0.944, -0.169)& 0.0202& *\\
2006-2008 vs 2010-2015& -0.6112& (-0.907, 0.089)& 0.0804& \\
2007-2008 vs 2010-2015& -0.6498& (-0.918, 0.025)& 0.0582& \\
\hline
\end{tabular}

\end{center}
\label{Table2}
\end{table}

\section{Discussion}
To our knowledge, this is the first study on the spatio-temporal patterns of influenza B proportion after the 2009 influenza pandemic, at both global and regional levels. We found an evidently positive correlation between the influenza B proportion and effective flight distance. We also found a significantly negative correlation  in the influenza B proportion between pre- and post-pandemic periods at the regional level in the US.

For the global patterns, we found that the influenza B proportion significantly correlated with the effective distance from Mexico, with a $p-$value $<0.0001$. This correlation is robust to the threshold of 500 cases. With higher threshold of 2000 cases, the correlation is still statistically significant and correlation coefficient increased. Also, excluding data from the year 2015 does not change the conclusion (Pearson's correlation=0.669, 95\% CI:0.422, 0.823, and p-value $<0.0001$). Again, it is hard to believe that these results are a coincidence.

Previous studies focused on the impacts of latitude on influenza B patterns. Baumgartner et al. studied the influenza activity worldwide and found its seasonality, timing of influenza epidemic period displayed unique patterns according to the latitudinal gradient \cite{Baumgartner}. Yu et al. studied the relationship between latitudinal gradient and influenza B proportion among the provinces in China. They have found an increasing prevalence of influenza B towards the South \cite{Yu}. Since influenza virus transmission varies with climatic factors such as temperature and humidity \cite{Lowen}, these influenza B patterns are likely to be explained by climatic factors that vary with the latitudinal gradients. However, at the global level, we did not find any evidence supporting the impact of latitude. But we observed an interesting relationship between influenza B proportion and longitude, which could be associated with the initial patterns of spread of H1N1pdm.

We found that the influenza B proportion in the period we considered is the lowest among the countries in the South America, followed by North America, Western Europe, Eastern Europe, Middle East and are the highest in Asia. We can see that European countries had a higher proportion of influenza B than United States. It is worthwhile to note that influenza B viruses are more likely to be reported from children \cite{Thompson}. Higher influenza vaccination coverage rate among the general population, and in particular, among healthy school children in the US and Canada, is likely to decrease the influenza B proportion in these two countries \cite{He+15}. It will be very worthwhile to conduct an in-depth study on these relationships in the future.

Our study found that there is a negative correlation between influenza B proportion and pre- and post-pandemic periods in the US. This could be due to cross-reactive immunity as suggested by Ahmed et al. where the A(H1N1) pandemic influenza resulted in a large number of persons exposed to this strain in certain countries could have resulted in anti-HA immunity \cite{Ahmed}. Whether this phenomenon is also applicable to influenza B virus is also worth further investigation.

The major strength of our study is the use of FluNet database, which allowed us to extract influenza data globally over a long period of time for a comprehensive analysis, and to establish the relationship between influenza B proportion and longitude and effective flight distance from Mexico respectively. Second, our use of longitude and effective flight distance in the study of influenza B proportion are both novel and epidemiologically relevant. Third, influenza B proportion is a more robust indicator of the severity of influenza B in a country than the total number of influenza B confirmations or influenza B positive rates (i.e. cases of influenza B out of all influenza specimens).

Our study had some limitations. The methods of surveillance data collection and their testing policies differ widely across different countries. The intensity of surveillance might change over time. Therefore our results should be interpreted with caution.

\section{Conclusions}

In this study, we have examined the spatio-temporal patterns of influenza B proportion globally and in the US. The impacts of different surveillance policies would have been reduced to some extent due to the use of proportion rather than other absolute numbers.

Our results showed that there displayed wide variations in the proportions of influenza B over the study period. Differences in influenza B proportion between Europe and Northern America (i.e. the US and Canada) could be associated with the different influenza vaccination policies and coverage among healthy school children and the general population. The patterns of spread of H1N1pdm could be one of the major reasons.  Future studies could examine whether other additional factors, e.g. health access rates, population age structure and climatic factors, which could contribute to the patterns of influenza B proportions. To identify the major factors for prioritizing public health control measures will be both challenging and worthwhile. The study is of both scientific and public health significance.

\section{Acknowledgments}
We are grateful for helpful discussions and suggestions from Ben Cowling, Lin Wang, Joe Wu and Lewi Stone. We are grateful to Lin Wang and Joe Wu for providing effective distance data. D.H. was supported by a RGC/ECS grant from Hong Kong Research Grant Council (25100114), and a Health and Medical Research Grant from Hong Kong Food and Health Bureau Research Council (13121382).

\section*{Contributions}

A.C., Q.L., D.Y. and D.H. conceived the work, analysed the data and wrote the manuscript. All authors reviewed the manuscript.

\section*{Competing interests}

The authors declare no competing financial interests.
\end{document}